\documentclass[article,twocolumn,showpacs,floatfix,longbibliography]{revtex4-2}
\usepackage{chngcntr}
\usepackage{mathrsfs,braket}
\usepackage{multirow}
\usepackage{amssymb, amsbsy, amsmath, latexsym, dsfont, array, layout, graphicx, mathrsfs, color, ulem, bm}
\usepackage[colorlinks=true,linkcolor=blue,anchorcolor=red,citecolor=blue, urlcolor=blue]{hyperref}

\begin{document}
\title{Lyapunov formulation of band theory for disordered non-Hermitian systems}
\author{Konghao Sun\textsuperscript{1}}
\author{Haiping Hu\textsuperscript{1, 2}}\email{hhu@iphy.ac.cn}
\affiliation{\textsuperscript{1}Beijing National Laboratory for Condensed Matter Physics, Institute of Physics, Chinese Academy of Sciences, Beijing 100190, China}
\affiliation{\textsuperscript{2}School of Physical Sciences, University of Chinese Academy of Sciences, Beijing 100049, China}
\begin{abstract}
Non-Bloch band theory serves as a cornerstone for understanding intriguing non-Hermitian phenomena, such as the skin effect and extreme spectral sensitivity to boundary conditions. Yet this theory hinges on translational symmetry and thus breaks down in disordered systems. Here, we develop a real-space Lyapunov formulation of band theory that governs the spectra and eigenstates of disordered non-Hermitian systems. This framework yields universal non-Hermitian Thouless relations linking spectral density and localization to Lyapunov exponents under different boundary conditions. We further identify an exact topological criterion: skin modes and Anderson-localized modes correspond to nonzero and zero winding numbers, respectively, revealing the topological nature of the skin-Anderson transition. This transition is dictated by an essential Lyapunov exponent and gives rise to novel unidirectional critical states. Our formulation provides a unified and exact description of spectra and localization in generic one-dimensional non-Hermitian systems without translational symmetry, offering new insights into the interplay among non-Hermiticity, disorder, and topology.
\end{abstract}
\maketitle
Non-Hermitian systems exhibit fundamentally distinct behaviors from their Hermitian counterparts \cite{coll1, coll4, coll6, colladd3, nhreview, nhreview2}. A paradigmatic example is the non-Hermitian skin effect (NHSE) \cite{nhse1,nhse2,nhse3,nhse4,nhse5,nhse6}, where a macroscopic number of eigenstates accumulate at system boundaries, as observed across a variety of experimental platforms \cite{exp1,exp2,exp3,exp4,exp5,exp6}. Non-Bloch band theory \cite{nhse1,nhse4,nhse5} provides a fundamental framework for understanding these peculiar phenomena. In particular, the NHSE originates from the presence of a point gap \cite{point1,point2,point3}, a feature unique to the complex energy spectra of non-Hermitian Hamiltonians. The extreme boundary sensitivity of both the spectrum and eigenstates goes beyond the conventional bulk–edge correspondence \cite{nhse7}, necessitating an analytical continuation of the lattice momentum $k\rightarrow\beta$. The allowed $\beta$ values define the so-called generalized Brillouin zone \cite{nhse1}, which captures the boundary localization of skin modes.

The non-Bloch band theory builds upon translational symmetry and breaks down when this symmetry is absent, for instance, in systems with disorder, external potentials, or applied fields. Notably, disorder competes with the NHSE by promoting bulk Anderson localization. Its interplay with non-Hermiticity has led to a variety of intriguing phenomena \cite{nhat1,nhat2,nhat3,nhat_syu,nhat4,nhat5,nhat6,nhat7,nhat8,nhat9,me0,me3,me5,me6,me7,me8,me9,taylor,wz_green,impurity1,impurity2,impurity3,impurity4,impurity5,impurity6,impurity7,lch,photonic_exp,wz_jump,hh_jump,hh_jump2}, e.g., new universality classes of Anderson transition \cite{nhat1,nhat2}, scale-free localization \cite{impurity1,impurity2,impurity3,impurity4,impurity5,impurity6,impurity7}, jumpy dynamics \cite{photonic_exp}, and universal wave transport \cite{wz_jump,hh_jump,hh_jump2}. These phenomena are ultimately rooted in the distinct spectral structure and localization behavior unique to non-Hermitian systems. To date, a unified framework governing energy spectra and eigenstates in generic non-Hermitian systems without translational symmetry remains elusive. This raises key questions: How can spectra and eigenstates be determined in disordered systems in the thermodynamic limit (TDL)? As disorder competes with the NHSE and drives a skin–Anderson transition, what is the universal feature of this transition? Are there any novel critical states emerging at the transition? Is there a topological criterion distinguishing skin modes from Anderson-localized modes (ALMs)?

In this work, we develop a unified non-Hermitian band theory for one-dimensional disordered lattices based on a real-space transfer matrix approach. We establish universal non-Hermitian Thouless relations that yield the spectral density and eigenstate localization under different boundary conditions in the TDL. We show that the transition between skin modes and ALMs is governed by an essential Lyapunov exponent (LE), $\gamma_{\mathrm{ess}}$, and can be tracked via the evolution of mobility edges in the complex energy plane. At the transition, we identify a novel type of unidirectional critical states (UCS), exhibiting localization in one direction and delocalization in the other. The spectral sensitivity to boundary conditions persists up to the onset of the Anderson insulating phase. We further establish an exact topological criterion: skin modes correspond to nonzero winding numbers, whereas ALMs correspond to zero winding number, revealing the topological nature of the skin–Anderson transition.

\begin{figure}[!t]
\centering
\includegraphics[width=3.33 in]{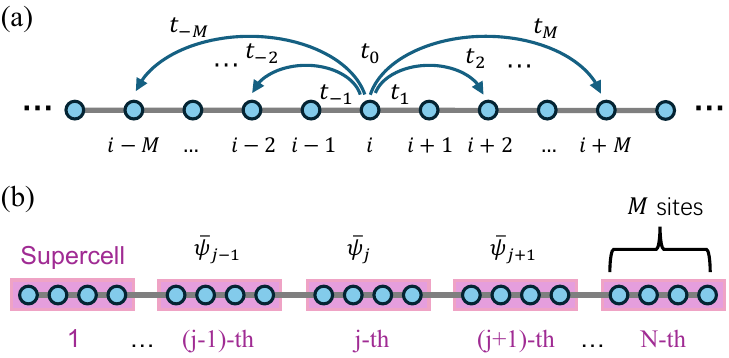}
\caption{Schematics of our real-space formulation for disordered lattices. (a) Sketch of a one-dimensional lattice with maximum hopping range $M$. Arrows indicate hoppings from a representative site. (b) The lattice is partitioned into a sequence of supercells (magenta blocks), each containing $M$ sites. The total system size is $L=NM$, with $N$ the number of supercells. The transfer matrix is constructed at the supercell level.}\label{fig1}
\end{figure}

{\bf Real-space formulation.}~We consider a generic one-dimensional disordered lattice with Hamiltonian
\begin{eqnarray}\label{model}
H=\sum_{|i-j|\leq M} t_{i,j}c_{i}^{\dag} c_j,
\end{eqnarray}
where $c_i$ is the annihilation operator at site $i$. $t_{i,j}$ is the hopping strength from site $i$ to $j$ and may take matrix values for systems with internal degrees of freedom (e.g., spin or sublattice). The system size is $L$ and the hopping range is $M$, as depicted in Fig. \ref{fig1}(a). The disorder can take any form, encoded in the entries $t_{i,j}$. For brevity, we label $t_{j-i}=t_{i,j}$ and treat it as a random variable whose distribution depends on the disorder type. For instance, Anderson-type onsite disorder corresponds to the diagonal element $t_0$ drawn from a uniform distribution. 

Disorder reshapes both the energy spectrum and eigenstate. On the spectral side, while clean non-Hermitian systems feature arc- or loop-shaped spectra under OBC or periodic boundary condition (PBC) \cite{graph,hu_PRB}, respectively, disorder typically broadens them into finite spectral areas in the complex plane. As the system size grows, the discrete eigenvalues become denser and approach a continuous distribution in the TDL. To determine the spectral density, we adopt an electrostatic analogy \cite{hu_PRB,potential1,hdnb2} and denote $E_n$ ($n=1,2,\dots,L$) as eigenenergies of the lattice Hamiltonian (\ref{model}). Treating each eigenvalue as a point charge of $1/L$ in the complex energy plane, the electrostatic potential at a point $E$ is
\begin{eqnarray}\label{col_pot}
\phi(E)=\frac{1}{L}\sum_{n=1}^L\ln|E_n-E|.
\end{eqnarray} 
In the TDL, the spectral density is given by the electrostatic potential via Poisson's equation [See Methods]. 

Disorder breaks translational symmetry, rendering lattice momentum ill-defined and invalidating the analytical continuation performed in clean systems. To analyze eigenstate localization (whether skin or Anderson), we introduce the supercell transfer matrix. The whole lattice is divided into a sequence of supercells, each containing $M$ sites, as shown in Fig. \ref{fig1}(b). The system size is set to be $L=NM$, with $N$ the number of supercells. In this notation, the Hamiltonian Eq. (\ref{model}) becomes block-tridiagonal [See Methods], with diagonal blocks describing intra-supercell couplings and off-diagonal blocks representing inter-supercell couplings. An eigenstate is expressed as $|\psi\rangle=(\bar{\psi}_1,\bar{\psi}_2,\cdots,\bar{\psi}_N)^T$, where each $\bar{\psi}_j$ ($j$ labels the supercell) is an $M$-component vector. Using these blocks, the eigenvalue equation $H|\psi\rangle=E|\psi\rangle$ is recast into transfer matrix form:
\begin{eqnarray}
  \left(\begin{array}{cc}\bar{\psi}_{j+1}\\
  \bar{\psi}_j\\
  \end{array}\right)= T_j(E) \left(\begin{array}{cc}
  \bar{\psi}_j\\
  \bar{\psi}_{j-1}\\
  \end{array}
  \right).
\end{eqnarray}
The $2M\times 2M$ matrix $T_j(E)$ depends on the energy $E$ and shifts the supercell by one. The boundary conditions are specified by $\bar{\psi}_{N+1}=\bar{\psi}_0=\bar{0}$ for OBC and $\bar{\psi}_{0}=\bar{\psi}_N$ for PBC. By successively applying these transfer matrices, the state is propagated from the leftmost to the rightmost supercell. The full transfer matrix is defined as
\begin{eqnarray}\label{t_matrix}
T(E) = \prod_{j=1}^{N} T_j(E).
\end{eqnarray}
According to Oseledec's ergodic theorem \cite{ergodic}, the eigenvalues of $\lim_{L\rightarrow\infty}\frac{1}{2L}\ln[T^{\dag}(E)T(E)]$, denoted as $e^{\gamma_j(E)}$ ($j=1,2,\cdots,2M$), yield $2M$ LEs in the TDL. These exponents, $\gamma_j(E)$, characterize the spatial decay of eigenstates and are arranged in ascending order: $\gamma_j(E)\leq\gamma_{j+1}(E)$. 

We now examine the connection between the spectrum and eigenstates in disordered lattices. As our first key result, we establish a direct relation between the spectral density and the LEs. In the TDL, We prove that the electrostatic potential [defined in Eq. (\ref{col_pot})] under OBC (PBC) relates to these LEs through [See Appendix \ref{S1} for OBC and Appendix \ref{S2} for PBC]:
\begin{eqnarray} 
\lim_{L\rightarrow\infty}\phi_{OBC}(E)&=&\sum_{s=M+1}^{2M}\gamma_s(E)+\mathbb{E}[\ln|t_M|];\label{pot_obc}\\
\lim_{L\rightarrow\infty}\phi_{PBC}(E)&=&\sum_{\gamma_s(E)>0}\gamma_s(E)+\mathbb{E}[\ln|t_M|].\label{pot_pbc}
\end{eqnarray}
Here, $\mathbb{E}[\cdot]$ means the ensemble average over disorder realizations. Remarkably, for OBC/PBC, the potential is determined by the \textit{$M$ largest} or \textit{positive} LEs, respectively. Accordingly, the spectral density under the two boundary conditions is given by Poisson's equation as
\begin{eqnarray}
\rho_{OBC}(E)&=&\frac{1}{2\pi}\sum_{s=M+1}^{2M}\nabla^2 \gamma_s(E);\label{den_obc}\\
\rho_{PBC}(E)&=&\frac{1}{2\pi}\sum_{\gamma_s(E)>0}\nabla^2 \gamma_s(E).\label{den_pbc}
\end{eqnarray}

We emphasize that Eqs. (\ref{den_obc})(\ref{den_pbc}) hold for systems with \textit{any finite-range} couplings and arbitrary disorder types. Beyond disordered systems, our framework can be extended to include external potentials and applied fields. These equations thus serve as universal relations for complex energy spectra, linking spectral density under different boundary conditions to LEs. We dub them non-Hermitian Thouless relations. In the clean limit, they reproduce the non-Bloch band theory and the generalized Brillouin zone [See Appendix \ref{S3}]. For the simplest case $M=1$ with only nearest-neighbor couplings, a related Thouless relation for OBC was previously derived for a specific class of off-diagonal disorder \cite{gtr}.

\begin{figure}[!t]
\centering
\includegraphics[width=3.33 in]{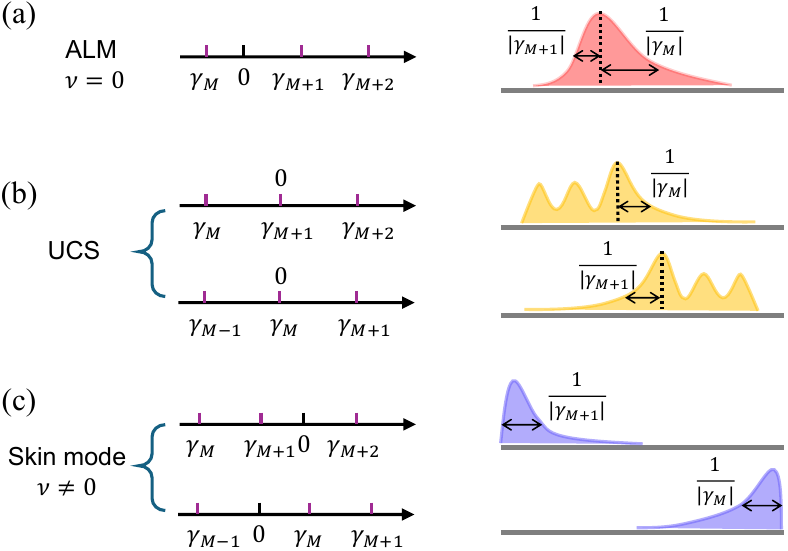}
\caption{Configurations of Lyapunov exponents (LEs) and their corresponding eigenstate profiles. (a) For an Anderson localized mode (ALM) in the bulk, the two central LEs, $\gamma_M$ and $\gamma_{M+1}$, lie on opposite sides of zero, resulting in direction-dependent localization lengths. (b) For a unidirectional critical state (UCS), either $\gamma_M=0$ or $\gamma_{M+1}=0$, indicating delocalization in one direction and localization in the other.  (c) For a skin mode, both central LEs lie on the same side of zero. The cases $\gamma_M < \gamma_{M+1} < 0$ and $0<\gamma_M < \gamma_{M+1}$ correspond to left and right skin modes, respectively. The winding number $\nu$ [see definition in Eq. (\ref{winding})] vanishes for ALMs and is nonzero for skin modes.}\label{fig2}
\end{figure}

{\bf Essential Lyapunov exponent.}~Based on the non-Hermitian Thouless relations, we now characterize the eigenstate localization. Generally, spectral sensitivity to boundary conditions stems from the presence of skin modes. Comparing Eqs. (\ref{den_obc}) and (\ref{den_pbc}), it is clear that the two central LEs, $\gamma_M$ and $\gamma_{M+1}$ play a key role. In the clean limit, for energies inside the OBC spectrum, they satisfy $\gamma_M(E)=\gamma_{M+1}(E)$, which defines the generalized Brillouin zone. In disordered systems, however, they generally differ, leading to two scenarios, as sketched in Figs. \ref{fig2}(a)(c). (i) $\gamma_M(E)<0<\gamma_{M+1}(E)$; (ii) both $\gamma_M(E)$ and $\gamma_{M+1}(E)$ are either negative or positive. In case (i), the Thouless relations imply $\rho_{OBC}(E)=\rho_{PBC}(E)$, indicating an ALM, while in (ii), $\rho_{OBC}(E)\neq\rho_{PBC}(E)$, signaling a skin mode. 

Figure \ref{fig2}(a) and \ref{fig2}(c) show representative spatial profiles for ALMs and skin modes, respectively. In non-Hermitian settings, transport can be direction-dependent due to nonreciprocity \cite{nhat_syu}. For ALMs, the two central LEs lie on opposite sides of zero, governing eigenstates' spatial decay in different directions. Specifically, the decay length from left to right is $1/|\gamma_M|$, while that in the opposite direction is $1/|\gamma_{M+1}|$, as determined by the inverse transfer matrix whose LEs are ordered as $-\gamma_{2M}\leq\cdots\leq -\gamma_{M+1}\leq-\gamma_M\leq\cdots$. For skin modes, when both central LEs are either negative or positive, the eigenstate accumulates at the left or right boundary, respectively. The localization length is then set by the more weakly decaying component ($1/|\gamma_M|$ v.s. $1/|\gamma_{M+1}|$), as shown in Fig. \ref{fig2}(c). Intriguingly, at the transition, either $\gamma_M$ or $\gamma_{M+1}$ crosses zero, as illustrated in Fig. \ref{fig2}(b). This gives rise to a novel type of critical mode with a hybrid character: it is delocalized in one direction while localized in the opposite direction. We term this the unidirectional critical state (UCS).

Now consider starting from the clean limit, where the two central LEs coincide. As disorder increases, they begin to separate, and one eventually crosses zero to the opposite side, signaling the transition from a skin mode to an ALM. To capture this, we define the essential Lyapunov exponent as the one closer to zero:
\begin{eqnarray}\label{ess}
\gamma_{ess}(E)= \begin{cases}
\gamma_M(E) & \text{if } |\gamma_M(E)| \leq |\gamma_{M+1}(E)|; \\
\gamma_{M+1}(E) & \text{if } |\gamma_M(E)|>|\gamma_{M+1}(E)|.
\end{cases}
\end{eqnarray}
The essential LE dictates the skin-Anderson transition, and the mobility edge satisfies: 
\begin{eqnarray}\label{mobility}
\gamma_{ess}(E)=0. 
\end{eqnarray}
This mobility edge typically forms closed contours in the complex energy plane \cite{footnote1}, separating regions of skin modes from those of ALMs. This analysis, grounded on the Thouless relations, offers a unified way to determine localization and pinpoint transitions between different types of modes—marking the second key result of our work.

\begin{figure}[!t]
\centering
\includegraphics[width=3.33 in]{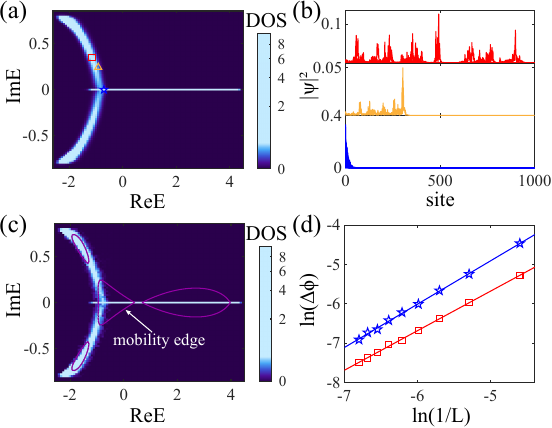}
\caption{Spectral density and eigenstate profiles under open boundary condition (OBC). (a) Spectral density from exact diagonalization, with average over $3200$ disorder realizations. The system size is $L=1000$. (b) Spatial profiles of eigenstates, selected from three small regions marked by red square, orange triangle, and blue star in (a) (from top to bottom). (c) Spectral density and mobility edge (magenta) determined by our theory [Eqs. (\ref{den_obc}) and (\ref{mobility})]. (d) Finite-size scaling of the potential deviation $\Delta\phi$ for $E=-0.6$ (blue star) and $E=-1.05+0.32i$ (red square), fitted by $\Delta\phi=e^{-0.635}L^{-1.008}$ (red line) and $\Delta\phi=e^{0.617}L^{-1.104}$ (blue line). The model is Eq. (\ref{model}) with $M=2$ and real onsite disorder. $W=0.8$, $t_2=0.5$, $t_1=1.5$, $t_{-1}=1$, $t_{-2}=1$.}\label{fig3}
\end{figure}
{\bf Illustrative examples.}~We present a concrete example with $M=2$ and real onsite disorder, where $t_{0}$ is uniformly drawn from $[-W,W]$. Figure \ref{fig3}(a) shows the spectral density at disorder strength $W=0.8$ by diagonalizing the OBC Hamiltonian and averaging over multiple disorder realizations. The theoretical prediction from Eq. (\ref{den_obc}) is shown in Fig. \ref{fig3}(c), closely matching the numerical results. To quantify the deviation, we examine the potential difference $\Delta\phi=|\phi-\phi_{OBC}|$, where $\phi$ is the electrostatic potential of the numerical spectra [Eq. (\ref{col_pot})] and $\phi_{OBC}$ is from Eq. (\ref{pot_obc}). A Finite-size analysis in Fig. \ref{fig3}(d) shows that $\Delta\phi$ vanishes algebraically in the TDL, confirming that the numerical spectrum converges to our theory. We also determine the mobility edge via the essential LE in Eq. (\ref{mobility}). The mobility edge forms closed contours [magenta curves in Fig. \ref{fig3}(c)] in the complex plane. Figure \ref{fig3}(b) displays the spatial profiles of eigenstates selected from small regions [marked by colored symbols in \ref{fig3}(a)], lying outside, on, and inside the contours, respectively. From top to bottom, they correspond to ALMs (ten eigenstates shown), hybrid critical mode, and skin mode. The hybrid mode features delocalization on one side and localization on the other, in sharp contrast to the multifractal states \cite{multifractal1} typical of Hermitian systems.

\begin{figure}[!t]
\centering
\includegraphics[width=3.33 in]{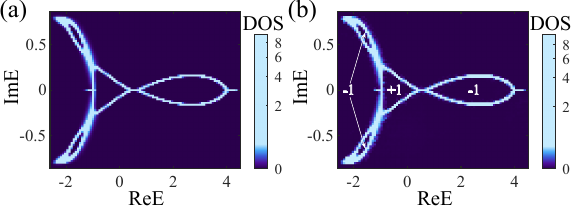}
\caption{Spectral density under periodic boundary condition (PBC) and winding-number criterion. (a) Spectral density obtained from exact diagonalization of PBC Hamiltonians. (b) Spectral density predicted by our formulation in Eq. (\ref{den_pbc}). The winding number is labeled for each spectral hole. Model parameters are the same as Fig. \ref{fig3}.}\label{fig4}
\end{figure}
For PBC, Figs.\ref{fig4}(a) and \ref{fig4}(b) show the spectral density from numerical diagonalization and our theory [via Eq. (\ref{den_pbc})], respectively, with excellent agreement. They show marked differences from the OBC case in Figs. \ref{fig3}(a)(c). Spectral sensitivity to boundary conditions occurs inside the mobility-edge contours: regions outside the contours (hosting ALMs under OBC) remain unaffected under PBC, while regions hosting skin modes under OBC become hollow under PBC, with the ``lost” states pushed onto and pinned to the contours. As a result, skin modes (and hybrid critical modes) delocalize under PBC. This intricate spectral sensitivity is \textit{universal} for disordered non-Hermitian systems and, as we will show, has a topological origin.

Through this example, we see that inferring the spectrum in the TDL via exact diagonalization typically requires large system sizes due to the slow algebraic convergence, and thus inevitably suffers from numerical errors. Our formalism, however, offers a robust way to achieve the same goal via LEs from $2M\times 2M$ matrices. We also studied other types of disorder [See Appendix \ref{S4}], such as off-diagonal and quasi-periodic cases, and found excellent agreement between our theory and numerical results. Notably, for the unidirectional hopping model (with only $t_1$ or $t_{-1}$ nonzero) with onsite disorder drawn from a distribution $\rho_w$, our theory yields $\rho_{OBC}=\rho_w$.

{\bf Topological criterion.}~The emergence of spectral holes under PBC is reminiscent of the point gap unique to complex spectra. In clean systems, skin modes originate from the point gap and are governed by a nontrivial winding number of the Bloch Hamiltonian \cite{point1,point2,point3}. In disordered systems, the lattice momentum is no longer a good quantum number. Here, we introduce a twisted boundary condition by setting $z=e^{i\theta}, \theta\in[0,2\pi]$ in the Hamiltonian (\ref{ham}). Physically, this corresponds to connecting the leftmost and rightmost supercells with an additional phase in the coupling. The winding number with respect to $E$ is defined as
\begin{eqnarray}\label{winding}
\nu(E)=\frac{1}{2\pi}\int_0^{2\pi}\frac{d}{d\theta}\arg\det[E-H(e^{i\theta})] d\theta.
\end{eqnarray}
We prove that $\nu(E)$ is given by the number of positive LEs (denoted $n_P(E)$) via [See Appendix \ref{S5}]
\begin{eqnarray}\label{winding2}
\nu(E) = M - n_P(E).
\end{eqnarray}
According to the Thouless relations in Eqs. (\ref{den_obc}) and (\ref{den_pbc}), we have
\begin{eqnarray}\label{criterion}
\begin{cases}
\rho_{OBC}(E)=\rho_{PBC}(E) & \text{if } \nu(E)=0; \\
\rho_{OBC}(E)\neq\rho_{PBC}(E) & \text{if } \nu(E)\neq 0.
\end{cases}
\end{eqnarray}
Noting that ALMs are insensitive to boundary conditions, we therefore establish an exact topological criterion: skin modes and ALMs correspond to nonzero and zero winding numbers, respectively. This constitutes the third key result of our work. Figure \ref{fig4}(b) confirms this correspondence: only the spectral holes (hosting skin modes under OBC), corresponding to skin modes under OBC, exhibit nonzero winding numbers.

\begin{figure}[!t]
\centering
\includegraphics[width=3.33 in]{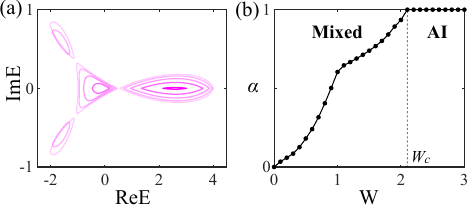}
\caption{Skin-Anderson transition tracked by mobility edges. (a) Mobility-edge contours in the complex plane for disorder strengths $W=0.4;0.8;1.2;1.6;2.0$. As $W$ increases (darker shades), the contours shrink. (c) The fraction of ALM $\alpha$ versus $W$. The threshold value is $W_c\approx 2.1$: below it, the system is in a mixed phase with both skin modes and ALMs; above it, the system enters an Anderson insulator (AI) phase hosting only ALMs. The model is the same as Fig. \ref{fig3}.}\label{fig5}
\end{figure}
{\bf Skin-Anderson transition.}~The competition between disorder and the NHSE drives the skin-Anderson transition. Our theory captures this transition and provides a typical picture of the entire process. Intuitively, different skin modes exhibit varying resilience to disorder. As the disorder strength $W$ increases, skin modes gradually morph into ALMs, accompanied by the essential LE $\gamma_{ess}(E)$ crossing zero. The system enters an Anderson insulator phase (featuring only ALMs) once $W$ exceeds a critical threshold $W_c$. Below $W_c$, skin modes and ALMs coexist, separated by mobility edges. The sensitivity to boundary conditions persists until the the transition is complete. This transition is tracked by the evolution of mobility edges and quantified by the ratio
\begin{eqnarray}
\alpha=\int_{\gamma_M(E)<0<\gamma_{M+1}(E)} \rho_{OBC}(E) dE,
\end{eqnarray}
which measures the fraction of ALMs in all eigenstates. In Fig. \ref{fig5}(a), we show the evolution of mobility-edge contours for various disorder strengths. Eigenmodes outside/inside these contours correspond to ALMs and skin modes, respectively. As $W$ increases, the contours shrink and eventually disappear, signaling the onset of the Anderson insulator. We also plot $\alpha$ versus $W$ in Fig. \ref{fig5}(b). The threshold value is determined to be $W_c \approx 2.1$, where $\alpha$ saturates to unity.

{\bf Discussion}

To summarize, we develop a unified non-Hermitian band theory for one-dimensional disordered systems based on a supercell transfer matrix approach. Built upon the non-Hermitian Thouless relations in Eqs. (\ref{den_obc}) and (\ref{den_pbc}), our framework accurately captures the spectral density, eigenstate localization, and mobility edges in the TDL through the LEs. Furthermore, we establish an exact correspondence between winding numbers and eigenmodes, and identify the skin–Anderson transition via the essential LE and evolution of mobility edges. Table \ref{table1} provides a side-by-side comparison between our Lyapunov formulation and the non-Bloch theory, and summarizes the main spectral and eigenstate results obtained from our theory.
\begin{table}[!t]
\centering
\begin{tabular}{|c|c|c|}
\hline
\textbf{Band theory} & \textbf{non-Bloch} & \textbf{Lyapunov} \\ \hline
Applicability & clean system & clean/disordered system  \\ \hline
Physical space & $k$-space & real space \\ \hline
\multirow{2}{*}{Spectrum} & PBC: $H(k)$~loop & $\rho_{PBC}\sim\sum_{\gamma_s>0}\nabla^2\gamma_s$\\ 
                           & OBC: $H(\beta)$~arc & $\rho_{OBC}\sim\sum_{s>M}\nabla^2\gamma_s$ \\ \hline
\multirow{2}{*}{Eigenstate} & PBC: Bloch & PBC: ALM/delocalized \\ 
                           & OBC: skin & OBC: ALM/UCS/skin\\ \hline
GBZ/LE & $\gamma_{M}=\gamma_{M+1}$ & $\gamma_{M}\neq\gamma_{M+1}$ \\ \hline
Mobility edge & $\times$ & $\gamma_{ess}(E)=0$ \\ \hline
Origin of NHSE & point gap & point gap \\ \hline
\end{tabular}
\caption{Comparison between our Lyapunov formulation and the non-Bloch band theory. Each row summarizes the applicable systems, physical operation space, energy spectrum, eigenstate types, generalized Brillouin zone (GBZ) condition or Lyapunov exponent (LE), mobility edge, and the origin of skin effect. For disordered systems, typical eigenstates under OBC include ALMs, skin modes, and unidirectional critical states (UCS) located on the mobility edge [see Fig. \ref{fig3}(b)]. Disorder breaks translational symmetry and the GBZ condition no longer holds. In the table, $\gamma_{ess}$ denotes the essential LE defined in Eq. (\ref{ess}).}\label{table1}
\end{table}

Our analysis extends to certain open quantum systems with disorder or random dissipation, described by the Lindblad master equation, as well as to random networks \cite{network1, network2}. The framework also applies to systems with spatial inhomogeneity such as impurities, external potentials, or fields \cite{impurity1,impurity2,impurity3,impurity4,impurity5,impurity6,impurity7,lch}. It offers new insights into state engineering, functionality design, and disorder-induced dynamical phenomena. For instance, recent studies have linked wave transport to spectral density near the imaginary band tail \cite{wz_jump, hh_jump, hh_jump2}, which can be scrutinized using our non-Hermitian Thouless relations. An open question remains how to extend the approach to higher dimensions, given the complexity of skin modes \cite{point2,hdnb1,hdnb3,hdnb4,hdnb5,hdnb7,hdnb9} and the diversity of lattice geometries. Experimentally, our predictions, such as the coexistence of skin modes and ALMs, their distinct sensitivity to boundary conditions, and the emergence of UCS, should be testable in synthetic platforms like quantum walks \cite{exp1}, topolectrical circuits \cite{exp2,exp3,exp4}, mechanical metamaterials \cite{exp5}, and cold atoms \cite{exp6}.\\

{\large{\bf Methods}}

{\bf Electrostatic analogy.}~The energy spectrum of a disordered non-Hermitian Hamiltonian consists of a set of discrete points in the complex plane $\mathbb{C}$. For an $L \times L$ Hamiltonian, denote its eigenvalues by $E_n$ ($n = 1, 2, \dots, L$). The spectral density, in the distributional sense, is given by:
\[
\rho(E) = \frac{1}{L} \sum_{n=1}^L \delta(E - E_n),
\]
where $\delta$ is the delta function. We are interested in the continuous distribution of eigenvalues in the thermodynamic limit (TDL) $L \to \infty$. To analyze this, we treat each eigenvalue as a point charge of magnitude $1/L$ in the complex plane. The electrostatic potential at a point $E\in\mathbb{C}$ is
\[
\phi(E) = \frac{1}{L} \sum_{n=1}^L \ln |E - E_n|.
\]
In the TDL, the electrostatic potential is determined by the spectral density through $\lim_{L\rightarrow\infty}\phi(E)=\int_{\mathbb{C}} \ln|E-E'|\rho(E')dE'$. Note that the Green's function $G(z) = \ln |z|$ is the fundamental solution to the two-dimensional Laplace equation, $\nabla^2 G(z) = 2\pi \delta(z)$. Thus, the spectral density is related to the potential via Poisson's equation:
\[
\nabla^2 \Phi(E) = 2\pi \rho(E).
\]

{\bf Block representation.}~The matrix form of the $L\times L$ disordered Hamiltonian in Eq. (\ref{model}) is
\begin{align}
	\left(\begin{array}{cccccc}
		t_{1,1} & t_{1,2} & \cdots  & t_{1,M+1}  & 0 &\\
		t_{2,1} & t_{2,2} &\cdots  & t_{2,M+1}  & t_{2,M+2} &\\
		\vdots &\vdots & \ddots &\vdots  & \vdots &\ddots\\	 
t_{M+1,1}& t_{M+1,2} & \cdots & t_{M+1,M+1} & t_{M+1,M+2} & \ddots\\
0  & t_{M+2,2} & \cdots & t_{M+2,M+1} &t_{M+2,M+2}&\ddots\\
 &  & \ddots &\ddots & \ddots & \ddots\\
	\end{array}\right).
	\label{model_text}
\end{align}
Here $M$ is the longest hopping range. The system length is set to be $L=MN$ ($N$ is an integer). In our transfer matrix approach, we divide the lattice into $N$ supercells, each containing $M$ lattice sites. For notational simplicity, we define $t_{j-i}=t_{i,j}$, where $t_{s}~(s=-M,-M+1,\cdots,M)$ are disordered parameters. 

To incorporate different boundary conditions (open/periodic/twisted boundary condition), we introduce a complex parameter $z$ and express the tight-binding Hamiltonian (\ref{model_text}) in a block-tridiagonal form:
\begin{eqnarray}\label{ham}
H(z)=\left(\begin{array}{ccccc}
h_1 & B_1 &      &  & z^{-1} C_1 \\
C_2 & h_2 & B_2 &   &  \\
 & C_3 & \ddots &\ddots &  \\
  & & \ddots &  & B_{N-1} \\
z B_N &    &  & C_N & h_N \\
\end{array}\right).
\end{eqnarray}
It is an $L\times L$ matrix with each entry being an $M\times M$ block. The diagonal blocks $h_j$ describe intra-supercell couplings, while the off-diagonal blocks $B_j$ and $C_j$ represent couplings between neighboring supercells. They take the form:
\begin{align}
h=\left(\begin{array}{cccc}
	t_0 & t_1  & \cdots & t_{M-1} \\
	t_{-1} & t_0  & \cdots & t_{M-2} \\
	\vdots & \vdots & \ddots & \vdots \\
	t_{-M+1} & t_{-M+2} & \cdots & t_0
\end{array}\right) ;
\end{align}
\begin{align}
C=\left(\begin{array}{cccc}
	t_{-M} & t_{-M+1} & \cdots & t_{-1} \\
	 & t_{-M} & \dots & t_{-2} \\
	  &  & \ddots & \vdots \\
	 &  &   & t_{-M}
\end{array}\right) ; 
\end{align}
\begin{align}
B=\left(\begin{array}{cccc}
	t_M &  &  &  \\
	t_{M-1} & t_M &  &  \\
	\vdots  & \vdots & \ddots &  \\
	t_1 & t_2 & \cdots & t_M
\end{array}\right).
\end{align}
$B_j$ and $C_j$ ($j=1,2,\cdots,N$) are lower and upper triangular. The OBC (PBC) is specified by the absence (presence) of corner blocks $C_1$ and $B_N$. For the twisted boundary condition, we set $z=e^{i\theta}$, $\theta\in[0,2\pi]$.

We represent the eigenvalue problem via the transfer matrix. To this end, let us take an eigenstate $\psi$ of the Hamiltonian (\ref{model_text}) and write it as $\psi=(\bar{\psi}_1,\bar{\psi}_2,\cdots,\bar{\psi}_N)$, which consists of $N$ supercells, each being an $M$-component vector. Using the supercell blocks, the eigenvalue equation $H|\psi\rangle=E|\psi\rangle$ reads
\begin{eqnarray}\label{eigeneq}
C_j\bar{\psi}_{j-1}+B_j\bar{\psi}_{j+1}+h_j\bar{\psi}_j=E\bar{\psi}_j,~(j=1,2,\cdots ,N).\notag\\
\end{eqnarray}
The boundary condition is specified by $\bar{\psi}_{N+1}=\bar{\psi}_0=\bar{0}$ for OBC and $\bar{\psi}_{0}=\bar{\psi}_N$ for PBC. Formally, Eq. (\ref{eigeneq}) can be recast into a transfer matrix form
\begin{eqnarray}
  \left(\begin{array}{cc}\bar{\psi}_{j+1}\\
  \bar{\psi}_j\\
  \end{array}\right)= T_j(E) \left(\begin{array}{cc}
  \bar{\psi}_j\\
  \bar{\psi}_{j-1}\\
  \end{array}
  \right),
\end{eqnarray}
with
\begin{eqnarray}\label{t_matrix1}
T_j (E)= \left(\begin{array}{cc}
B_j^{-1} (E - h_j) & -B_j^{-1} C_j \\
I_{M\times M} & 0 \\
\end{array}\right)
\end{eqnarray}
a $2M\times 2M$ matrix. $T_j(E)$ depends on the choice of $E$. In one single step, the transfer matrix shifts the supercell by one. By successively applying the transfer matrices, the state component at the leftmost supercell is linked to that at the rightmost supercell. The full transfer matrix is defined as Eq. (\ref{t_matrix}) in the main text.\\



{\bf Acknowledgements}

This work is supported by the National Key Research and Development Program of China (Grants No. 2023YFA1406704 and No. 2022YFA1405800) and National Natural Science Foundation of China (Grant No. 12474496).\\

\clearpage
\appendix
\setcounter{equation}{0}  
\setcounter{figure}{0}
\counterwithout{equation}{section}  
\renewcommand{\thefigure}{S\arabic{figure}}
\renewcommand{\thesection}{S\arabic{section}}
\pagebreak
\renewcommand{\theequation}{S\arabic{equation}}
\widetext
\begin{center}
\textbf{\large  Supplemental Material for ``Lyapunov formulation of band theory for disordered non-Hermitian systems"}
\end{center}

\begin{center}
\vspace{0.5em}
Konghao Sun\textsuperscript{1} and Haiping Hu\textsuperscript{1, 2}\email{hhu@iphy.ac.cn}

\vspace{0.3em}
\textsuperscript{1}Beijing National Laboratory for Condensed Matter Physics, Institute of Physics, Chinese Academy of Sciences, Beijing 100190, China\\
\textsuperscript{2}School of Physical Sciences, University of Chinese Academy of Sciences, Beijing 100049, China
\end{center}
\vspace{3\baselineskip}

This Supplemental material (SM) provides additional details on
	
(S1) Derivation of the electrostatic potential [Eq. (\ref{pot_obc}) in the main text] under open boundary condition (OBC);

(S2) Derivation of the electrostatic potential [Eq. (\ref{pot_pbc}) in the main text] under periodic boundary condition (PBC);

(S3) Reduction to the clean case;
	
(S4) More examples with different types of disorder;

(S5) Proof of Eq. (\ref{winding2}) in the main text.

\section{Derivation of the electrostatic potential under OBC}\label{S1}
In this section, we derive the potential function [Eq. (\ref{pot_obc}) in the main text] for disordered non-Hermitian systems under OBC. We begin by expressing the potential function [defined in Eq. (\ref{col_pot}) of the main text] in terms of the characteristic polynomial of the OBC Hamiltonian:
\begin{eqnarray}\label{rel4}
\phi_{OBC}(E)=\frac{1}{L}\sum_{n=1}^L\ln\left|E_n-E\right|=\frac{1}{L}\ln\left|\det[H_{OBC}-E]\right|.
\end{eqnarray}
Here $E_n$ ($n=1,2,\dots,L$) is $n$-th eigenvalue of $H_{OBC}$. To compute the characteristic polynomial, we use the block-tridiagonal form of $H_{OBC}$ in Eq. (\ref{ham}). Applying transfer matrix technique, the eigenvalue equation $H_{OBC}|\psi\rangle=E|\psi\rangle$ is recast into the following equations:
\begin{eqnarray}
(h_1-E)\bar{\psi}_1+B_1\bar{\psi}_2&=&0;\label{beq1}\\
B_j\bar{\psi}_{j+1}+(h_j-E)\bar{\psi}_j+C_{j-1}\bar{\psi}_{j-1}&=&0;~~~~(1<j<N)\\
(h_N-E)\bar{\psi}_N+C_{N-1}\bar{\psi}_{N-1}&=&0.\label{beq2}
\end{eqnarray}
Here the eigenstate $\psi=(\bar{\psi}_1,\bar{\psi}_2,\cdots,\bar{\psi}_N)$ is split into $N$ segments, each being an $M$-component vector. The boundary condition is set by the constraint $\bar{\psi}_{N+1}=\bar{\psi}_0=\bar{0}$. In the bulk, the one-step supercell transfer matrix $T_j~(1<j<N)$ is defined as
\begin{eqnarray}
T_j (E)= \left(\begin{array}{cc}
B_j^{-1} (E - h_j) & -B_j^{-1} C_j \\
I_{M\times M} & 0 \\
\end{array}\right).
\end{eqnarray}
It shifts the supercell one by one, i.e., $(\bar{\psi}_{j+1},\bar{\psi}_j) = T_j(E) (\bar{\psi}_j, \bar{\psi}_{j-1})$ for $1<j<N$. For the two boundary equations [Eqs. (\ref{beq1})(\ref{beq2})], we define two boundary matrices
\begin{eqnarray}
T_1^b (E)= \left(\begin{array}{cc}
B_1^{-1} (E - h_j) & -B_1^{-1} \\
I_{M\times M} & 0 \\
\end{array}\right);~~~~T_N^b (E)= \left(\begin{array}{cc}
(E  - h_N) & -C_{N-1} \\
I_{M\times M} & 0 \\
\end{array}\right).
\end{eqnarray}
The full transfer matrix under OBC is then constructed as below: 
\begin{eqnarray}
T^{OBC}(E)=T_N^b(T_{N-1}T_{N-2}\cdots T_3T_2)T_1^b.
\end{eqnarray}
It is easy to check that $T^{OBC}(E)$ transfers the supercell from one end to the other end of the chain:
\begin{eqnarray}\label{eigeneq2}
\left(\begin{array}{c}
0\\
\bar{\psi}_N \\
\end{array}\right)=T^{OBC}(E)\left(\begin{array}{c}
\bar{\psi}_1 \\
0\\
\end{array}\right).
\end{eqnarray}

Next, we express $T^{OBC}(E)$ in terms of its blocks as below:
\begin{eqnarray}
T^{OBC} (E)= \left(\begin{array}{cc}
T^{OBC}_{(11)}(E) & T^{OBC}_{(12)}(E) \\
T^{OBC}_{(21)}(E) & T^{OBC}_{(22)}(E) \\
\end{array}\right),
\end{eqnarray}
where $T^{OBC}_{(11)}$, $T^{OBC}_{(12)}$, $T^{OBC}_{(21)}$, $T^{OBC}_{(22)}$ denote the upper-left, upper-right, lower-left, and lower-right corner of $T^{OBC}$ respectively. Each block has dimensions $M \times M$. From Eq. (\ref{eigeneq2}), it follows that $\det [T^{OBC}_{(11)}(E)]=0$ if $E$ is inside the OBC spectrum. Moreover, we can get $\det [H_{OBC}-E]=0$ from the eigenvalue equation $(H_{OBC}-E)|\psi\rangle=0$ for $E$ inside the spectrum. This indicates that $\det [H_{OBC}-E]=0$ and $\det[T^{OBC}_{(11)}(E)]=0$ share the same $L$ roots. Since both equations are polynomials in $E$ of degree $L$, we conclude that $\det[H_{OBC}-E]\propto\det [T^{OBC}_{(11)}(E)]$. To determine the exact proportionality coefficient, we analyze the large $E$ limit. In this limit, $\det[H_{OBC}-E]=(-1)^L E^L+O(E^{L-1})$ and $\det [T^{OBC}_{(11)}(E)]=\prod_{j=1}^N\det[B_j^{-1}]E^L+O(E^{L-1})$, we thus have the following relation:
\begin{eqnarray}\label{rel5}
\det [H_{OBC}-E]=(-1)^L\det[T^{OBC}_{(11)}(E)]\det[B_1 B_2\cdots B_{N-1}].
\end{eqnarray}
Furthermore, we can derive:
\begin{align}\label{rel6}
    \lim_{L\rightarrow\infty}\phi_{OBC}\left(E\right)=\lim_{L\rightarrow\infty}\frac{1}{L}\ln\left|\det\left[H_{OBC}-E\right]\right|=\lim_{L\rightarrow\infty}\frac{1}{L}\ln\left|\det\left[T^{OBC}_{\left(11\right)}\left(E\right)\right]\right|+\mathbb{E}[\ln|t_M|],
\end{align}
where $\mathbb{E}[\cdot]$ takes ensemble average over disorder realizations. 

We proceed to relate $\det [H_{OBC}-E]$ to the Lyapunov exponents (LEs). Let us scrutinize the structure of the full transfer matrix $T^{OBC} (E)$. First, we claim that it yields the same LEs as $T(E)$ [See its definition in Eq. (\ref{t_matrix}) of the main text] in the thermodynamic limit (TDL). In fact, the only difference between $T^{OBC}(E)$ and $T(E)$ arises at the boundaries. Specifically, $T(E)=T_N(T_N^b)^{-1}T^{OBC}(E)(T_1^b)^{-1}T_1$. The LEs are defined as the eigenvalues of the matrix $\lim_{L\rightarrow\infty}\frac{1}{2L}\ln[T^{\dag}(E)T(E)]$. Due to the $1/L$ factor, the LEs are irrelevant to these boundary terms in the TDL. 

Next, we perform a similarity transformation on $H_{OBC}$ as $SH_{OBC}S^{-1}$. Here $S$ is given by:
\begin{eqnarray}
	S=\left(\begin{array}{cccc}
        e^{M\xi}I_{M\times M} & & &\\
        & e^{2M\xi}I_{M\times M} & &\\
        & & \ddots &\\
        & & &  e^{NM\xi}I_{M\times M}
            \end{array}\right).
\end{eqnarray}
Under this transformation, $B_j\rightarrow e^{-M\xi}B_j$ and $C_j\rightarrow e^{M\xi}C_j$. Consequently, $T_j$ transforms as $T_j\rightarrow e^{M\xi} OT_j O^{-1}$, where:
\begin{eqnarray}
	O=\left(\begin{array}{cc}
		e^{M\xi/2}I_{M\times M} &\\
		& e^{-M\xi/2}I_{M\times M} 
	\end{array}\right).
\end{eqnarray}
As a result, the LEs shift as $\gamma_s\rightarrow\gamma_s+\xi$. On the other hand, Eq. (\ref{rel6}) remains invariant under this similarity transformation (The OBC spectrum stays intact under similarity transformation), but $\mathbb{E}[\ln|t_M|]$ transforms into $\mathbb{E}[\ln|t_M|]-M\xi$. It follows that $\lim_{L\rightarrow\infty}\frac{1}{L}\ln\left|\det\left[T^{OBC}_{\left(11\right)}\left(E\right)\right]\right|$ shifts as $ \lim_{L\rightarrow\infty}\frac{1}{L}\ln\left|\det\left[T^{OBC}_{\left(11\right)}\left(E\right)\right]\right|+M\xi$. Note that these shifts are independent of $E$. It implies that $\mathbb{E}[\ln|t_M|]$ equals the sum of some $M$ LEs (the $E$-irrelevant constant is unimportant for our discussion). Furthermore, in the clean limit, Eq. (\ref{clean}) tells that $\lim_{L\rightarrow\infty}\frac{1}{L}\ln\left|\det\left[T^{OBC}_{\left(11\right)}\left(E\right)\right]\right|$ corresponds to the sum of the largest $M$ LEs. In general case, the LEs vary continuously with disorder strength $W$. We thus conclude that in the presence of disorder, $\lim_{L\rightarrow\infty}\frac{1}{L}\ln\left|\det\left[T^{OBC}_{\left(11\right)}\left(E\right)\right]\right|$ is given by the sum of the largest $M$ LEs, namely:
\begin{eqnarray}\label{rel8}
	\lim_{L\rightarrow\infty}\frac{1}{L}\ln|\det[T^{OBC}_{(11)}(E)]|=\sum_{s=M+1}^{2M}\gamma_{s}(E).
\end{eqnarray}
By combining Eq. (\ref{rel6}) and (\ref{rel8}), we arrive at the final form of the potential function [see Eq. (\ref{pot_obc}) in the main text]:
\begin{eqnarray}
\lim_{L\rightarrow\infty}\phi_{OBC}(E)=\sum_{s=M+1}^{2M}\gamma_s(E)+\mathbb{E}[\ln|t_M|].
\end{eqnarray}

\section{Derivation of the electrostatic potential under PBC}\label{S2}
For the PBC case, we consider the following $L\times L$ Hamiltonian with parameter $z$:
\begin{align}
	H(z)=\left(\begin{array}{ccccccc}
		h_1 & B_1 & & & & C_1/z \\
		C_2 & h_2 & B_2 & & \\
		& C_3 & h_3 & B_3 &  &  \\
		&  & \ddots & \ddots & \ddots &  \\
		&  &  & C_{N-1} & h_{N-1} & B_{N-1} \\
		zB_N &  &  &  & C_{N} & h_N
	\end{array}\right)_{N \times N},
\end{align}
The case $z=1$ corresponds to the PBC Hamiltonian. Similar to the OBC case, we have the following expression of the potential function under PBC:
\begin{eqnarray}\label{rel7}
\phi_{PBC}(E)=\frac{1}{L}\sum_{n=1}^L\ln|E_n-E|=\frac{1}{L}\ln|\det[H_{PBC}-E]|.
\end{eqnarray}

An important result regarding such block tridiagonal matrices with corner blocks is the duality relation \cite{duality_ref} which relates the determinant of the transfer matrix to the determinant of $H(z)$:
\begin{align}\label{duality2}
\operatorname{det}[T(E)-z]=(-z)^M \frac{\operatorname{det}[E-H(z)]}{\operatorname{det}\left[B_1 B_2 \ldots B_N\right]}.
\end{align}
This duality relation immediately yields:
\begin{equation}
\begin{aligned}
	& \ln \left|\operatorname{det}\left[E-H\left(e^{L \xi+i \theta}\right)\right]\right|=\sum_{j=1}^N \ln \left|\operatorname{det}B_j\right|-M L\xi+\frac{1}{2} \sum_{s=1}^{2M} \ln \left|e^{L\gamma_s+i \theta_s}-e^{L \xi+i  \theta}\right|^2 \\
	& =\sum_{j=1}^N \ln \left|\operatorname{det}B_j\right|+\frac{1}{2} \sum_{s=1}^{2M}\left[L\gamma_s+\ln \left(2 \cosh \left(L \gamma_s-L \xi\right)-2 \cos \left(\theta_s- \theta\right)\right)\right],
\end{aligned}
\end{equation}
with $\xi= \frac{\ln \left|z\right|}{L}$ and $\theta=\arg z$. From the identity: $\ln |2 \cosh x-2 \cos y|=|x|-2 \sum_{l=1}^{\infty} \frac{\cos l y}{l} e^{-l |x|}$, it follows that:
\begin{align}
\ln \left|\operatorname{det}\left[E-H\left(e^{L\xi+i \theta}\right)\right]\right|=\sum_{j=1}^N \ln \left|\operatorname{det}B_j\right|+\frac{L}{2} \sum_{s=1}^{2M} \gamma_s+\frac{L}{2} \sum_{s=1}^{2M} \left|\gamma_s-\xi\right|-\sum_{l=1}^{\infty} \sum_{s=1}^{2M} \frac{\cos l \left(\theta_s-\theta\right)}{l} e^{-l L \left|\gamma_s-\xi\right|} .
\label{det_pbc}
\end{align}
The PBC corresponds to the $z=1$ case. In the TDL, $L\rightarrow\infty$, the last term in the above equation is negligible, yielding:
\begin{align}
	\phi_{P B C}\left(E\right)=\frac{1}{L}\ln\left|\operatorname{det}\left[E-H\left(z=1\right)\right]\right|
    =\sum_{\gamma_s > 0} \gamma_s+\mathbb{E}\left[\ln \left|t_M\right|\right],
\end{align}
where $\mathbb{E}[\cdot]$ denotes ensemble average over disorder realizations. Note the difference from the OBC case. For OBC and PBC, the summation is performed over the largest $M$ and positive LEs, respectively.

\section{Reduction to the clean case}\label{S3}
In this section, we demonstrate that our formalism naturally recovers the non-Bloch band theory developed through analytical continuation, when the disorder strength is zero. We begin with a brief overview of the non-Bloch band theory in clean non-Hermitian systems. An alternative description of the non-Bloch bands utilizes the potential function, which offers more information (e.g., spectral density), beyond the generalized Brillouin zone (GBZ). We then derive the potential function for the non-Bloch bands. Finally, we show how our Lyapunov formulation for disordered systems recovers the non-Bloch band theory when the disorder strength is zero.

\subsection{Recap of the non-Bloch band theory in clean systems}
For the considered systems with hopping-range $M$, the Bloch Hamiltonian is $H(k)=\sum_{s=-M}^M t_s e^{i s k}$. By analytical continuation of the lattice momentum $k\rightarrow \beta=e^{i k}$, we have the characteristic polynomial:
\begin{eqnarray}\label{chp1}
	f(\beta,E)=\sum_{s=-M}^M t_s \beta^s-E.
\end{eqnarray}
It has two complex variables. For a given $E$, there exists $2M$ roots  $\{\beta_1(E),\beta_2(E),...,\beta_{2M}(E)\}$ satisfying $f(\beta,E)=0$. Using these roots, the characteristic polynomial Eq. (\ref{chp1}) is represented as
\begin{eqnarray}
	f(\beta,E)=\frac{t_M}{\beta^M}\prod_{s=1}^{2M}(\beta-\beta_s(E)).
\end{eqnarray}
We arrange these roots in ascending order of their moduli, $|\beta_1(E)|\leq |\beta_2(E)|\leq\cdots\leq \beta_{2M}(E)$. In the thermodynamic limit (TDL), the OBC spectra (i.e., the non-Bloch spectra) are given by those $E$ satisfying
\begin{eqnarray}
	|\beta_{M}(E)|=|\beta_{M+1}(E)|.
\end{eqnarray}
The trajectories of these $\beta$-roots form the GBZ. The GBZ forms closed contours in the complex-$\beta$ plane. The non-Bloch spectra form arcs in the complex energy plane.

\subsection{Potential function for the clean case}
In electrostatic analog, each eigenenegy of the Hamiltonian is treated as a charged particle. In the TDL, the Coulomb potential contributed by the non-Bloch spectra is given by
\begin{eqnarray}\label{pot_1}
	\lim_{L\rightarrow \infty}\phi_{OBC}(E)=\oint_{GBZ}\frac{d\beta}{2\pi i\beta} \ln |f(\beta,E)|.
\end{eqnarray}
By inserting Eq. (\ref{chp1}), we have
\begin{eqnarray}\label{pot_1}
	\lim_{L\rightarrow \infty}\phi_{OBC}(E)=\ln |t_M|+\oint_{GBZ}\frac{d\beta}{2\pi i\beta}(-M\ln|\beta|+\sum_{s=1}^{2M}\ln|\beta-\beta_s|).
\end{eqnarray}
We split the sum over $s$ into two parts, $1\le s\le M$ and $M+1\le s\le 2M$. For $E$ outside the non-Bloch spectrum, $|\beta_{M}(E)|\neq|\beta_{M+1}(E)|$. The GBZ forms a closed loop around the origin, enclosing exactly $M$ $\beta$-roots inside and the other $M$ roots outside. Therefore, the contour integral over the GBZ can be replaced by an integral over a circle of radius $R$, where $|\beta_{M}(E)|<R<|\beta_{M+1}(E)|$. We then have
\begin{eqnarray}\label{clean}
	\lim_{L\rightarrow \infty}\phi_{OBC}(E)&=&\ln|t_{M}|-M\ln|R|+\int_{0}^{2\pi}\frac{d\theta}{2\pi}\sum_{s=1}^{2M}\ln|Re^{i\theta}-\beta_{s}(E)|\notag\\
	&=&\ln|t_{M}|-M\ln|R|+\sum_{s=1}^{2M}\ln\max[R,\beta_{s}(E)]\notag\\
	&=&\ln|t_{M}|+\sum_{s=M+1}^{2M}\ln|\beta_{s}(E)|.\label{pot_2}
\end{eqnarray}
This reproduces the potential function for clean systems obtained in Ref. \cite{hu_PRB}. In the above derivation, we have set $E$ outside the non-Bloch spectrum. However, the expression can be extended to the entire complex plane. The spectral density is determined by the Poisson equation: $\rho_{OBC}(E)=\frac{1}{2\pi}\nabla^2\phi_{OBC}(E)$. It can be further shown that \cite{hu_PRB,hdnb2} $\rho_{OBC}(E)=0$ if $|\beta_M(E)|\neq|\beta_{M+1}(E)|$. Thus, the GBZ condition is obtained. The non-Bloch spectrum corresponds to those $E$ with $|\beta_M(E)|=|\beta_{M+1}(E)|$. Compared to the potential function for disordered systems [Eq. (\ref{pot_obc}) in the main text], the similarity is evident. In the disordered case, the constant term in Eq. (\ref{pot_2}) is replaced by it ensemble average. The summation over the $\beta$-roots is replaced by the summation over the Lyapunov exponents (LEs). Below, we show that the $\beta$ roots in the clean case indeed correspond to LEs in our transfer matrix formalism. 

\subsection{The LEs in the clean case}
We start from the eigenvalue equation $H_{OBC}|\psi\rangle=E|\psi\rangle$ with $\psi=\left(\psi_1,\psi_2,\cdots,\psi_L\right)$ an $L$-component vector. Explicitly, we write the eigenvalue equation in terms of the $2M\times 2M$ transfer matrix:

\begin{align}	
	\left(\begin{array}{c}
		\psi_{M+l} \\
		\psi_{M-1+l} \\
		\vdots \\
		\psi_{l} \\
		\psi_{-1+l} \\
		\vdots \\
		\psi_{-M+2+l} \\
		\psi_{-M+1+l}
	\end{array}\right)=\tilde{T}_l(E)\left(\begin{array}{c}
		\psi_{M-1+l} \\
		\psi_{M-2+l} \\
		\vdots \\
		\psi_{-1+l} \\
		\psi_{-2+l} \\
		\vdots \\
		\psi_{-M+1+l} \\
		\psi_{-M+l}
	\end{array}\right);~~\tilde{T}_l(E)=\left(\begin{array}{cccccccc}
		-\frac{t_{M-1}}{t_M} & -\frac{t_{M-2}}{t_M} & \cdots &  -\frac{t_1}{t_M} & \frac{E-t_0}{t_M} & -\frac{t_{-1}}{t_M}  & \cdots & -\frac{t_{-M}}{t_M} \\
		1 &  &   &  &  &  &  &  \\
		& 1 &  &  &  &  &  &  \\
		&  & \ddots &  &  &  &  &  \\
		&  &  & 1 &  &  &  &  \\
		&  &  &  & 1 &  &  &  \\
		&  &  &  &  & 1 &  &  \\
		&  &  &  &  &  & \ddots & 
	\end{array}\right).
\end{align}
This matrix has only non-zero elements in the subdiagonal and the first row. Compared to the supercell transfer matrix $T_j(E)$ [defined in Eq. (\ref{t_matrix1}) of the main text] which shifts a supercell (of $M$ lattice sites) at one step, the matrix $\tilde{T}_l$ shifts a single site in each step. Note that they correspond to the same eigenvalue problem and satisfy $T_j(E)=\prod_{m=1}^M \tilde{T}_{j M+m}(E)$. The full transfer matrix is defined as
\begin{align}
	\tilde{T}(E)=\prod_{l=1}^L \tilde{T}_l(E).
\end{align}
For the clean case, the one-step matrix $\tilde{T}_l(E)$ does not depend on $l$. Therefore, the LEs directly relates to the eigenvalues of $\tilde{T}_l(E)$. Due to its simple form, we have
\begin{eqnarray}
	\det(z-\tilde{T}_l(E))=\sum_{s=0}^{2M}\frac{t_{-M+s}}{t_M}z^s=\frac{z^M}{t_M}f(z,E).
\end{eqnarray}
Here $f(z,E)$ is nothing but the characteristic polynomial of the Hamiltonian in Eq. (\ref{chp1}). Thus, the LEs correspond to the roots of the characteristic polynomial as
\begin{eqnarray}
	\ln|\beta_{s}(E)|\rightarrow \gamma_s(E).
\end{eqnarray} 

\section{More examples with different types of disorder} \label{S4}
\begin{figure}[hbpt]
	\centering
	\includegraphics[scale=1]{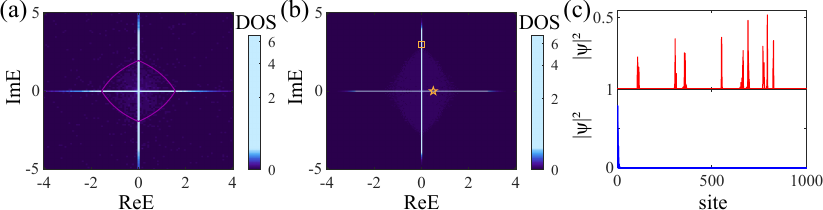}
	\caption{Spectral density and mobility edge for the Hatano-Nelson model with off-diagonal disorder. (a) Spectral density predicted by Eq. (\ref{den_obc}) in the main text, with the mobility edge (in magenta) determined by the condition $\gamma_{ess}(E)=0$. (b) Numerical spectral density obtained from exact diagonalization of the OBC Hamiltonian. The system size is $L=1000$, and average over $3200$ disorder realizations is taken. (c) Spatial profiles of ten eigenstates selected from two small regions on either side of the mobility edge. The upper (lower) panel corresponds to the square (star) marker in (b). The parameters are $t=1$, $\gamma=1.4$ and $W=2$. }\label{figs2}
\end{figure}

We provide additional examples to demonstrate the broad applicability of our theory. The first one is the Hatano-Nelson model with off-diagonal disorder:
\begin{align}
      H=\sum_{i=1}^{L-1} \left(t-\gamma+w_i\right) c^{\dagger}_{i+1} c_i+\left(t+\gamma+w_i\right) c^{\dagger}_i c_{i+1}.
\end{align}
Here $w_i$ represents the off-diagonal disorder, uniformly distributed in $\left[-W, W\right]$. We set $W=2$ and impose OBC. Figure \ref{figs2}(a) presents the spectral density predicted by our formulation in Eq. (\ref{den_obc}) of the main text. The mobility edge, determined by the condition $\gamma_{ess}(E)=0$, forms a closed loop that separates the skin modes (inside the loop) from the Anderson localized modes (ALMs) (outside the loop). This agrees well with the spectral density obtained from exact diagonalization of the OBC Hamiltonian, as shown in Fig. \ref{figs2}(b). We select representative eigenstates from inside and outside the mobility-edge loop and plot their spatial profiles in Fig. \ref{figs2}(c), confirming their correspondence to skin modes and ALMs, respectively.

\begin{figure}[hbpt]
	\centering
	\includegraphics[scale=1]{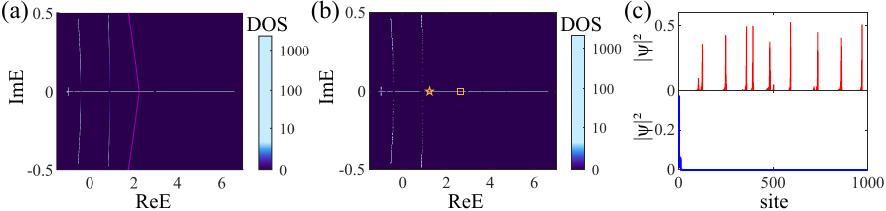}
	\caption{Spectral density and mobility edge in the Hatano-Nelson model with quasi-periodic disorder. (a) Spectral density predicted by Eq. (\ref{den_obc}) in the main text, with the mobility edge (in magenta) determined by the condition $\gamma_{ess}(E)=0$. (b)Spectral density obtained from exact diagonalization of the OBC Hamiltonian. The system size is $L=1000$. (c) Spatial profiles of ten eigenstates selected from two small regions (marked in (b)) on either side of the mobility edge. The upper (lower) panel corresponds to square (star) marker in (b). The model parameters are $t=1$, $\gamma=1.1$, $b=0.7$ and $\lambda=1$.}
	\label{figs3}
\end{figure}

The second one is the same model but with a quasi-periodic onsite potential. The Hamiltonian is
 \begin{align}
 	H=\sum_{i=1}^{L-1} \left(t-\gamma\right) c^{\dagger}_{i+1} c_i+\left(t+\gamma\right) c^{\dagger}_i c_{i+1}+\frac{2\lambda\cos\left(2\pi\omega i\right)}{1-b\cos\left(2\pi\omega i\right)} c^{\dagger}_i c_i,
 \end{align}
with $\omega=\frac{\sqrt{5}-1}{2}$. We set the strength of the quasi-periodic potential to $\lambda=1$ and impose OBC. Figure \ref{figs3}(a) shows the spectral density predicted by our theory along with the mobility edge locations, which closely match the spectral density from exact diagonalization in Fig. \ref{figs3}(b). Figure \ref{figs3}(c) plots the spatial profiles of eigenstates selected from both sides of the mobility edge, confirming their correspondence to ALMs and skin modes, respectively.

\section{Proof of Eq. (\ref{winding2}) in the main text}\label{S5}
The winding number is defined as:
\begin{align}
	\nu(E)=\frac{1}{2 \pi} \int_0^{2 \pi} \frac{d}{d \theta} \arg \det\left[E-H(e^{i \theta})\right] d \theta.
\end{align}
By employing the duality relation in Eq. (\ref{duality2}), we have
\begin{align}
	\det\left[E-H(e^{i \theta})\right]=(-1)^M \frac{1}{e^{i M \theta}} \det\left[B_1 B_2 \ldots B_N\right] \det\left[T(E)-e^{i \theta}\right] .
\end{align}
In the $L\to \infty$ limit, we have:
\begin{equation}
	\begin{aligned}
		\arg\det\left[E-H(e^{i \theta})\right]&=\mathrm{constant}-M \theta+2 M \theta+\sum_{s=1}^{2M} \arg \left[e^{L\gamma_s+i (\theta_s-\theta)}-1\right],\\
		&\approx \text { constant }+M\theta+ \sum_{\gamma_s>0}\left(\theta_s-\theta\right) .
	\end{aligned}
\end{equation}
It follows immediately that $\nu(E)=M-n_P(E)$, where $n_P(E)$ is the number of positive LEs among all. Therefore, $\nu(E)=0$ iff $n_P(E)=M$. For disordered systems, a non-vanishing winding number indicates the presence of skin modes under OBC.


\begin{thebibliography}{99}
\bibitem{footnote1} $\gamma_{ess}(E)$ can be regarded as a real function of complex variable $E$. A co-dimensional analysis of $\gamma_{ess}(E)=0$ indicates that  the mobility-edge form closed contours in the complex plane.
\end{thebibliography}
\end{document}